\documentclass[lettersize,journal]{IEEEtran}
\usepackage{amsmath,amsfonts}
\usepackage{algorithmic}
\usepackage{algorithm}
\usepackage{array}
\usepackage[caption=false,font=normalsize,labelfont=sf,textfont=sf]{subfig}
\usepackage{textcomp}
\usepackage{stfloats}
\usepackage{url}
\usepackage{verbatim}
\usepackage{graphicx}
\usepackage{cite}
\usepackage{booktabs}   
\usepackage{amssymb}    
\usepackage{amsmath}    
\usepackage{multirow}
\usepackage{graphicx}   

\usepackage{siunitx}
\usepackage{amsmath,amssymb,amsfonts}
\usepackage{wrapfig}
\usepackage{threeparttable}

\begin{document}

\title{SC-Match: Scale-Space Matching with Context Consistency for Side-Scan Sonar Mapping}

\author{Can Lei,
        Rafael Garcia,~\IEEEmembership{Member,~IEEE},
        Nuno Gracias,~\IEEEmembership{Member,~IEEE},
        Hayat Rajani,~\IEEEmembership{Member,~IEEE},    
        and Huigang Wang,~\IEEEmembership{Member,~IEEE}
\thanks{Can Lei and Huigang Wang are with the School of Marine Science and Technology, Northwestern Polytechnical University, Xi'an 710072, China. Huigang Wang is also with the Research \& Development Institute of Northwestern Polytechnical University in Shenzhen, Shenzhen, China. Rafael Garcia, Nuno Gracias and Hayat Rajani are with the Computer Vision and Robotics Research Institute (ViCOROB) of the University of Girona, Girona 17001, Spain. This work was conducted while Can Lei was on a research stay at ViCOROB, Spain. Corresponding author: Huigang Wang (e-mail: wanghg74@nwpu.edu.cn).}

\thanks{This work was partly supported by the Spanish government through projects ASSiST (PID2023-149413OB-I00) and IURBI (CNS2023-144688). This work was also supported by the Technology and Innovation of Shenzhen Municipality (JCYJ20241202124931042, ZDCYKCX20250901093900002).}}%
\maketitle

\begin{abstract}
Reliable estimation of spatial correspondences between overlapping side-scan sonar (SSS) measurements is essential for mapping, but acoustic appearance variations, weak seabed texture, repetitive patterns, and shadows make such correspondences sparse, unstable, and context-dependent. Scarce point-level annotations further limit sonar-specific training or fine-tuning of deep matching models. To this end, we propose SC-Match, a training-free scale-space matching framework with context-consistent correspondence refinement that adapts pretrained feature extraction and matching components to SSS observations without sonar-specific retraining. The framework improves correspondence reliability through scale-space representation and context-consistent refinement. For feature representation, a frozen extractor is applied to multiple observation scales, detector responses are calibrated according to the structure--texture tendency of the input SSS image, and score-aware cross-scale fusion is used to retain compact feature candidates. For correspondence refinement, adjacent local matching cases are used as neighboring contexts to verify stable fixed--moving relations and preserve non-conflicting complementary matches for alignment. Experiments on datasets acquired in different environments using different sonar platforms show that SC-Match provides more accurate correspondences and more consistent geometric alignment than representative pretrained methods, while maintaining stable behavior under unseen cross-platform acquisition conditions.

\end{abstract}


\section{Introduction}

\IEEEPARstart{S}{ide-scan} sonar (SSS) is widely used for seafloor mapping, marine archaeology, and benthic habitat observation \cite{A1}. By transmitting acoustic pulses and recording backscattered echoes from both sides of a moving platform, SSS systems efficiently acquire strip observations of large seabed areas \cite{A2}. In practical mapping applications, zigzag trajectories and intentionally overlapped tracks often provide repeated or partially overlapping observations of the same seabed region. These observations must be spatially related before they can be integrated into a geometrically consistent seabed representation. Therefore, reliable estimation of spatial correspondences between overlapping SSS observations is essential for mosaic construction and georeferenced map refinement.

However, estimating such correspondences is challenging. Unlike optical images, SSS observations are generated by acoustic backscattering and are jointly affected by grazing angle, propagation loss, sonar altitude, seabed relief, and platform attitude \cite{A3}. Consequently, the same seabed region may exhibit inconsistent intensity patterns and apparent structural variations across different survey passes. Weak seabed texture, repetitive ripple patterns, speckle-like noise, and elongated shadows further reduce feature repeatability and increase descriptor ambiguity \cite{A4}. These factors make correspondence evidence sparse, unstable, and dependent on the surrounding acoustic context, limiting the direct transfer of image matching methods developed primarily for optical imagery.

Existing methods generally estimate spatial relations between SSS observations using image-derived matching cues, external geometric priors, or learned correspondence models. Classical image-derived methods establish correspondences directly from sonar images \cite{A5}, but their reliability decreases under cross-track acoustic appearance variations, particularly in weak-texture, repetitive, and shadow-affected regions. Navigation- or geocoding-assisted methods provide useful overlap priors and initial geometric constraints \cite{A6}; however, residual positioning errors, platform-state uncertainty, and seabed-induced local distortions still require image-level correspondence refinement. Recent learned features and matchers provide stronger correspondence modeling \cite{A7}, but most pretrained models are developed using optical imagery, while sonar-specific retraining is restricted by the scarcity of reliable point-level correspondence annotations. Existing approaches are therefore limited by appearance-sensitive image evidence, approximate external priors, or labeled-data requirements that are difficult to satisfy in practical SSS surveys. Taken together, these limitations hinder the reliable extraction of cross-track spatial information from repeated acoustic measurements under sensing variability, limited navigation accuracy, and scarce point-level supervision.

To address these limitations, we propose SC-Match, a training-free scale-space matching framework with context-consistent correspondence refinement for SSS mapping. It adapts pretrained feature extraction and matching components to sonar observations at inference time without retraining or fine-tuning. The framework exploits two properties of SSS surveys. First, acoustic structures and seabed textures exhibit different saliency across observation scales, providing complementary correspondence evidence. Second, repeated survey observations contain recurrent representations of the same seabed regions, enabling correspondence reliability to be evaluated across neighboring local contexts. The main contributions are summarized as follows.

\begin{itemize}
    \item 
    We propose SC-Match, a training-free SSS matching framework that obtains reliable spatial correspondences between overlapping observations by exploiting scale-dependent acoustic saliency and recurrent local observations, without sonar-specific point-level supervision.

    \item     
    We introduce a content-adaptive scale-space feature generation strategy that calibrates multi-scale detector responses according to the structure--texture tendency of the input SSS observation and fuses scale-space candidates into a compact feature set containing complementary structural and textural cues.

    \item    
    We design a context-consistent correspondence generation and alignment strategy that uses adjacent local observation pairs to retain reproduced fixed--moving relations and preserve non-conflicting complementary correspondences for confidence-guided filtering and RANSAC-based geometric alignment.
\end{itemize}

\section{Related Work}

\subsection{Classical Image-Derived Matching}

Classical image-derived methods rely on manually designed cues, such as intensity, gradients, structures, or transform-domain coefficients \cite{A5}. These methods are training-free and can be roughly grouped into handcrafted-feature methods and transform-domain methods.

Handcrafted-feature methods estimate geometric relations from salient image structures. Wang et al. \cite{A9} combined spatial-gradient feature blocks with A-KAZE matching, RANSAC, and seam-line fusion to improve underwater terrain image stitching under feature-sparse conditions. Shang et al. \cite{A10} improved mismatch removal by combining feature-point clustering with motion analysis, reducing the dependence on a predefined global transformation model. Cui et al. \cite{A11} used phase congruency, edge and corner features, and hybrid descriptors for sonar image registration, improving robustness to grayscale differences. Although these methods improve feature extraction or mismatch filtering, their reliability is still limited by the repeatability of handcrafted saliency cues under changing acoustic appearances. Transform-domain methods represent sonar images in frequency or multiscale geometric domains to improve registration or fusion. Zhang et al. \cite{A12} used curvelet transform with resolution constraints to preserve clearer information in SSS strip mosaicking. Zhao et al. \cite{A13} further incorporated SSS-specific characteristics into curvelet-domain fusion to preserve complementary strip details and reduce shadow effects. These methods can improve post-registration image quality, but they usually rely on stable global correlations or transform-specific parameter settings.

Overall, classical image-derived methods provide practical non-learned solutions for SSS mapping, but they remain sensitive to acoustic appearance changes, weak or repetitive textures, shadow-induced structures, and locally varying distortions. This has motivated learning-based methods for more robust feature representation and correspondence modeling.

\begin{figure*}[htbp]
	\centering	\includegraphics[width=1\linewidth]{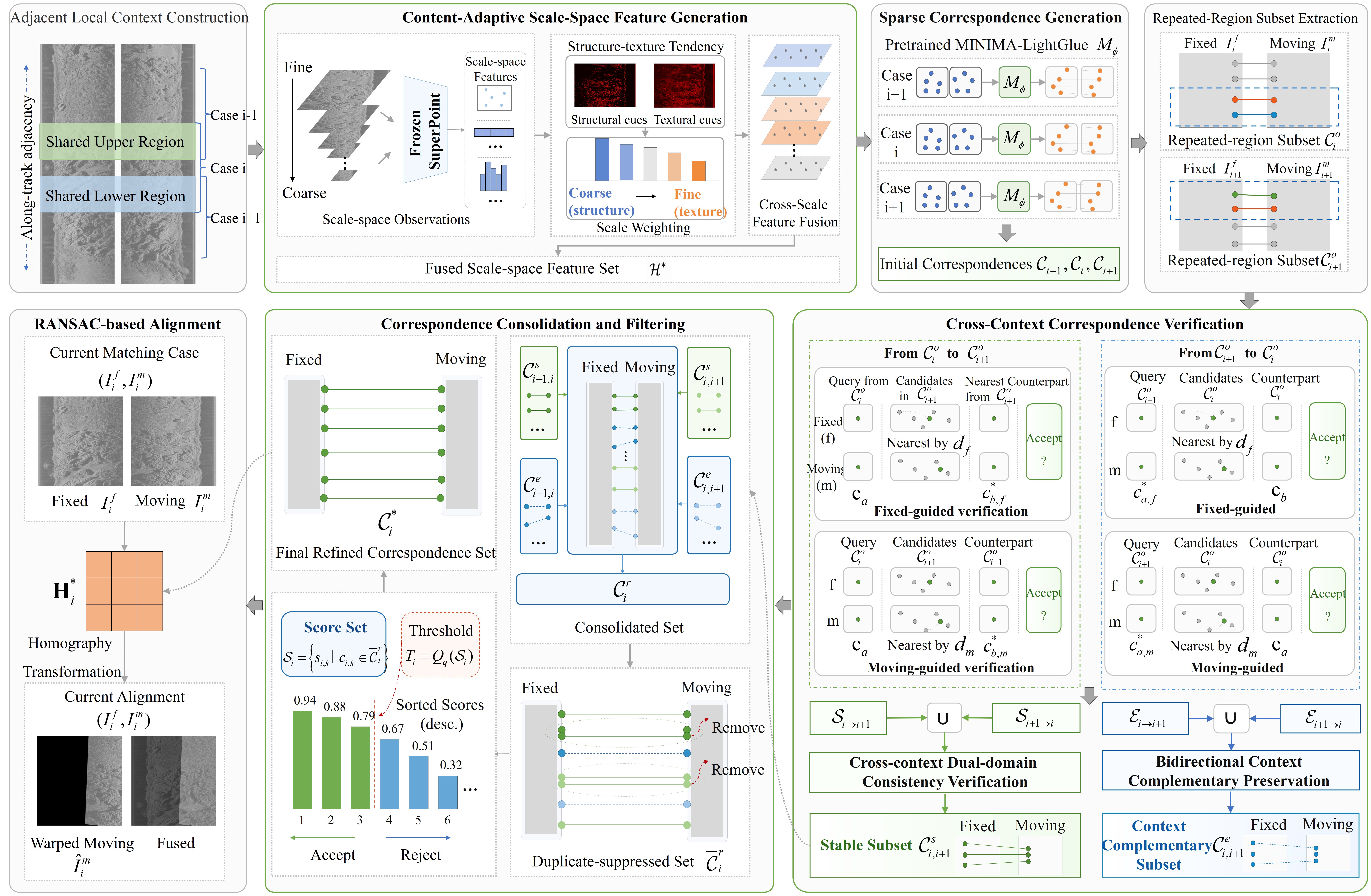}
    \caption{Overall framework of the proposed SC-Match method. Adjacent local fixed--moving cases are constructed with shared repeated regions along the track. For each local observation, a frozen SuperPoint extractor generates scale-space candidates, whose responses are calibrated by the structure--texture tendency and fused into $\mathcal{H}^{*}$ through score-aware cross-scale fusion. The fused features are matched by pretrained MINIMA-LightGlue, and repeated-region correspondences are extracted in a common coordinate frame. Cross-context dual-domain verification produces stable and context-complementary correspondences, which are consolidated, duplicate-suppressed, confidence-filtered, and used for RANSAC-based alignment.}
	\label{method01}
\end{figure*}	

\subsection{Navigation- and Geocoding-Assisted Methods}

Navigation- and geocoding-assisted methods use positioning measurements, track-line geometry, and platform pose information to provide geometric constraints for SSS mapping \cite{A6}. 

Zhao et al. \cite{A15} constructed a coarse SSS mosaic in a geographic frame using geocoded image coordinates and then used track-line constraints to maintain global spatial stability during local adjustment of overlapping regions. Shang et al. \cite{A16} further automated this geocoding-assisted workflow by determining overlapping areas from track lines and swath width, and by using geographic-coordinate constraints to guide local mosaic refinement. Zhang et al. \cite{A17} focused on AUV-collected SSS data, deriving the coordinate transformation between carrier and navigation frames to support autonomous frame-level stitching and handle gaps and overlaps during vehicle operation. In broader sonar mapping, Rypkema and Singh \cite{A18} introduced a hybrid LBL-iUSBL acoustic pose estimation system and fused acoustic pose estimates with multibeam scan matching, showing that improved acoustic positioning can enhance sonar mapping when conventional navigation is unreliable. Overall, navigation- and geocoding-assisted methods are effective for coarse spatial organization and geometric ambiguity reduction. However, navigation errors, attitude uncertainty, towfish motion, and interpolation errors can still limit their accuracy, so reliable image-level correspondences remain necessary for accurate local alignment.

\subsection{Learning-Based Methods}

Learning-based methods aim to improve feature representation and correspondence modeling through data-driven descriptors \cite{E1}, attention mechanisms \cite{E2}, or neural similarity estimation \cite{A7}. In sonar-related tasks, recent studies have introduced deep models for sonar image matching and mapping \cite{E3}. Yang et al. \cite{A20} proposed a SSS image matching method based on topological representation, where semantic segmentation, knowledge distillation, attention matrices, and graph neural networks were combined to learn image-level structural similarity. Hu et al. \cite{A21} addressed sonar image-based underwater terrain matching by learning spatial feature distribution transformations, improving robustness to noise, angular deviation, and locally similar terrain patterns. Zhang et al. \cite{A22} studied acoustic--optical joint underwater object detection and used cross-modality feature matching to handle weak alignment between sonar and optical observations. These studies show the potential of learning-based representation for underwater matching, but they usually require task-specific training data and are not directly designed for SSS mapping.

In the broader optical image matching field, learned methods can be roughly grouped into CNN-based and transformer-based approaches. Among CNN-based methods, XFeat \cite{A23} focuses on efficient feature extraction and supports both sparse and semi-dense matching with a lightweight architecture, while RIPE \cite{A24} learns robust keypoint detection and description from weakly supervised image pairs using reinforcement learning. Transformer-based methods further improve correspondence modeling through global or dense feature interaction. E-LoFTR \cite{A25} establishes correspondences through efficient dense feature interaction and improves matching in weak-texture regions, while RDD \cite{A26} uses a deformable-transformer-based descriptor branch to capture global context and geometric invariance, enabling robust matching under large viewpoint and scale variations. MINIMA \cite{A27} further attempts to reduce appearance discrepancies across imaging conditions or modalities. Although these methods provide powerful pretrained matching components, they are mainly developed on optical or non-sonar data, whose image statistics differ substantially from SSS observations.

Overall, learning-based methods provide stronger correspondence modeling than handcrafted operators, but their application to SSS mapping remains constrained by domain discrepancy and the scarcity of reliable point-level annotations. Sonar-specific training or fine-tuning is therefore difficult in practical mapping scenarios, whereas directly applying pretrained optical-image matchers may still be unreliable under acoustic appearance variations and local ambiguity. This motivates a training-free adaptation strategy that uses pretrained matching components while explicitly accounting for the acquisition characteristics of SSS data.

\section{Methodology}

\subsection{Overview}

Reliable SSS mapping requires robust feature representation and reliable correspondence support under scale-dependent acoustic responses, partial overlap, and local appearance ambiguity. Fig. \ref{method01} shows the overall workflow of the proposed SC-Match framework. Adjacent fixed--moving SSS observations are first organized into local matching cases with shared along-track regions. A scale-space feature generation module then produces fused feature sets for each observation, which are matched within each fixed--moving case to obtain initial correspondences. Repeated-region subsets from adjacent cases are compared across neighboring contexts to retain stable correspondences and preserve context-complementary support. The resulting correspondences are consolidated, duplicate-suppressed, confidence-filtered, and used for RANSAC-based homography estimation.

\begin{figure*}[htbp]
	\centering	\includegraphics[width=1\linewidth]{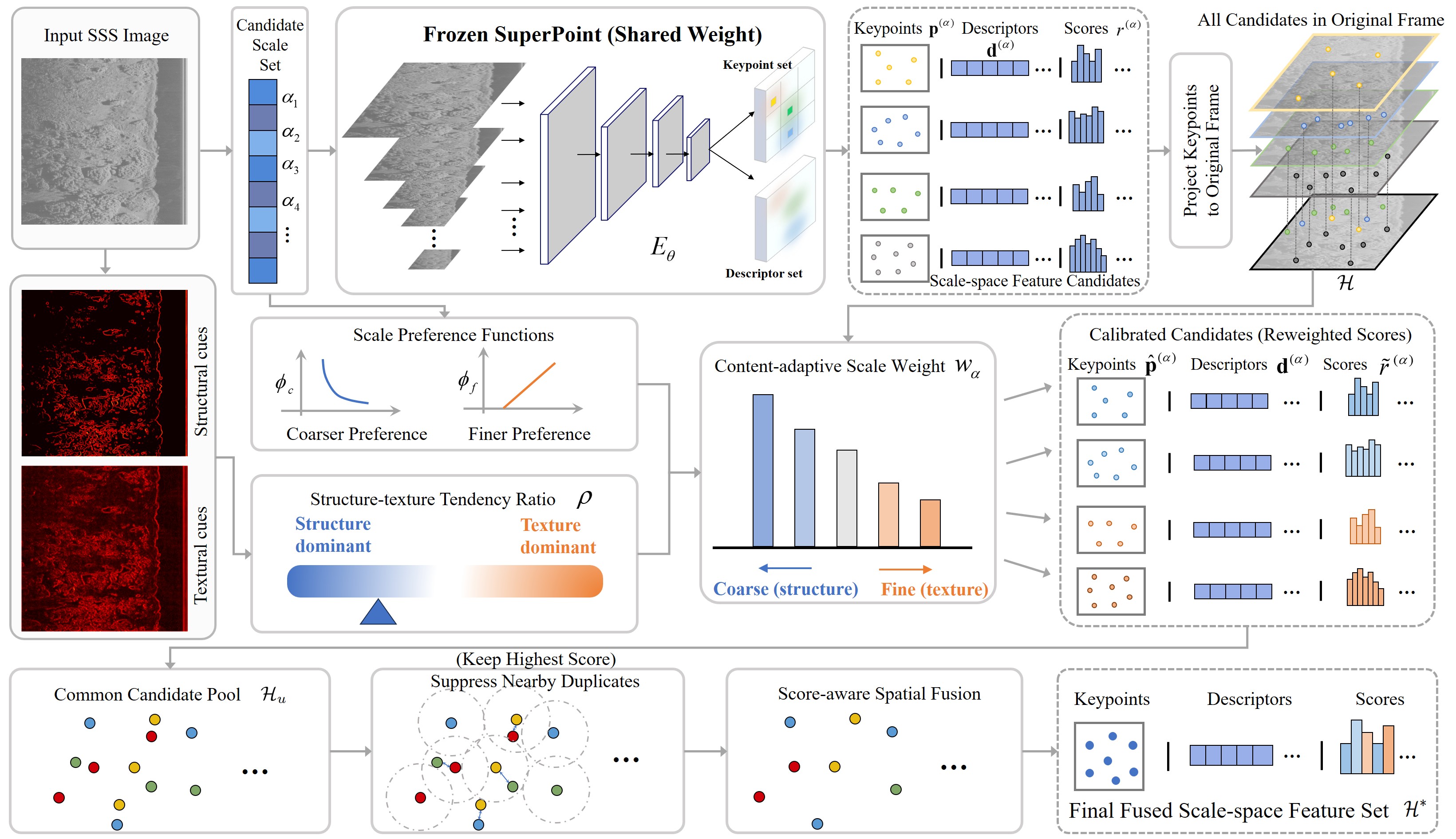}
    \caption{Scale-space feature generation. A candidate scale set is used to construct a scale-space observation pool, and a frozen SuperPoint extractor with shared weights generates keypoints, descriptors, and detector scores at each scale. After projecting keypoints back to the original image frame, structural and textural cues are used to estimate the structure--texture tendency ratio and compute content-adaptive scale weights from coarse- and fine-scale preference functions. The weights recalibrate detector scores without changing keypoint locations or descriptors. Finally, calibrated candidates from all scales are collected into a common candidate pool and integrated by score-aware spatial fusion, which produces the final fused scale-space feature set $\mathcal{H}^{*}$.}
	\label{method02}
\end{figure*}	

\subsection{Scale-Space Feature Generation}

Feature extraction in SSS imagery is highly sensitive to the observation scale, because acoustic boundaries, local scattering patterns, and seabed textures may not reach their maximum saliency at the same scale. Coarser observations tend to emphasize large acoustic structures such as object contours, seabed morphology, and shadow boundaries, whereas finer observations better preserve local scattering variations and texture details. Since reliable point-level correspondence annotations are scarce for sonar-specific fine-tuning, we adapt a frozen extractor at inference time through scale-space candidate generation, content-adaptive response calibration, and cross-scale feature fusion, as shown in Fig. \ref{method02}.

\subsubsection{Scale-Space Feature Extraction}

Given an input SSS image $I\in\mathbb{R}^{H\times W}$, a candidate scale set $\cal{A}$ is defined to construct a scale-space observation pool:
\begin{equation}
    {\cal I} = \left\{ {{I^{(\alpha )}}\mid \alpha  \in {\cal A}} \right\},\quad {I^{(\alpha )}} = {{\cal T}_\alpha }(I),
\end{equation}
where ${{\cal T}_\alpha }( \cdot )$ resizes the input image from $H\times W$ to $\alpha H\times \alpha W$. The same frozen SuperPoint extractor $E_{\theta}$ \cite{A28} is then applied separately to each observation $I^{(\alpha)}$ to generate a set of scale-space feature candidates:
\begin{equation}
    {{\cal H}^{(\alpha )}} = {E_\theta }\left( {{I^{(\alpha )}}} \right) = \left\{ {\left( {{\bf{p}}_i^{(\alpha )},{\bf{d}}_i^{(\alpha )},r_i^{(\alpha )}} \right)} \right\}_{i = 1}^{{N_\alpha }},
\end{equation}
where $\theta$ denotes the fixed pretrained parameters \cite{A28}, ${\bf{p}}_{i}^{(\alpha)}$ is the detected keypoint location in the $\alpha$-scale coordinate frame, ${\bf{d}}_{i}^{(\alpha)}$ is the corresponding descriptor, and $r_{i}^{(\alpha)}$ is the detector response score. To compare and fuse candidates from different scales, all detected keypoints are projected back to the original image coordinate frame:
$ \widehat{\bf{p}}_{i}^{(\alpha)}={\bf{p}}_i^{(\alpha )}/\alpha.$

\subsubsection{Content-Adaptive Response Calibration}

The scale-space candidates generated above are obtained from multiple observation scales, and their fusion priority is content-dependent. Here, acoustic content is defined as the observable composition of salient structural cues and fine-scale textural/scattering cues in the current SSS observation. This composition is quantified by the structure--texture tendency and mapped to scale-level response weights. These weights are used to calibrate detector scores so that cross-scale fusion can prioritize candidates whose scales are more consistent with the current SSS content.


Let $I_g$ denote the grayscale SSS image $I$ normalized to $[0,1]$. Two lightweight energy indicators are computed from $I_g$. The first one measures sparse but salient structural responses. Since such structures usually produce strong gradient responses, the horizontal and vertical Sobel gradients are denoted as $G_x$ and $G_y$, and the gradient magnitude at pixel location ${\bf{x}}=(x,y)^T$ is computed as
\begin{equation}
    G({\bf{x}}) = \sqrt {{G_x}{{({\bf{x}})}^2} + {G_y}{{({\bf{x}})}^2}} .
\end{equation}
The structural energy is then averaged over high-gradient locations:
\begin{equation}
    {E_s}({I_g}) = \frac{1}{{|{\Omega _s}|}}\sum\limits_{{\bf{x}} \in {\Omega _s}} G ({\bf{x}}),\quad {\Omega _s} = \left\{ {{\bf{x}}\mid G({\bf{x}}) > {\mu _G} + {\sigma _G}} \right\},
\end{equation}
where $\mu_G$ and $\sigma_G$ are the mean and standard deviation of $G(\bf{x})$, respectively.

The second textural indicator measures distributed fine-scale intensity variation and scattering fluctuation. It is computed over the whole image domain using local variance and Laplacian responses:
\begin{equation}
    {E_t}({I_g}) = \frac{1}{{|\Omega |}}\sum\limits_{{\bf{x}} \in \Omega } {\left[ {{{{\mathop{\rm Var}\nolimits} }_{{\cal N}({\bf{x}})}}\left( {{I_g}} \right) + \left| {{\nabla ^2} {I_g}({\bf{x}})} \right|} \right]} ,
\end{equation}
where $\rm{Var}_{\cal{N}(\bf{x})}(\cdot)$ denotes the local variance within the neighborhood $\cal{N}(\bf{x})$, and ${\nabla ^2}$ is the Laplacian operator. In our implementation, the local variance is computed using average pooling.

Based on these two indicators, the structure--texture tendency ratio is defined as
\begin{equation}
    \rho ({I_g}) = \frac{{{E_t}({I_g})}}{{{E_t}({I_g}) + {E_s}({I_g}) + \epsilon}},
\end{equation}
where $\epsilon$ is a small constant for numerical stability. $\rho ({I_g})$ should be interpreted as a relative content tendency within the current observation. A smaller $\rho(I_g)$ suggests stronger structural dominance, whereas a larger $\rho(I_g)$ suggests more prominent local texture and scattering variation.

To map this observation-level ratio to scale-level reliability weights, two scale preference functions are defined:
\begin{equation}
    {\phi _c}(\alpha ) = \max \left( {\frac{1}{\alpha } - 1,0} \right),\quad {\phi _f}(\alpha ) = \max \left( {\alpha  - 1,0} \right),
\end{equation}
where $\phi_c(\alpha)$ assigns a positive preference to coarser observations with $\alpha<1$, while $\phi_f(\alpha)$ assigns a positive preference to finer observations with $\alpha>1$. The native observation with $\alpha=1$ is treated as a neutral reference.

Accordingly, the content-adaptive scale weight is finally computed as
\begin{equation}
    {w_\alpha }({I_g}) = \frac{{\exp \left[ {(1 - \rho ({I_g})){\phi _c}(\alpha ) + \rho ({I_g}){\phi _f}(\alpha )} \right]}}{{\frac{1}{{|{\cal A}|}}\sum\limits_{\alpha ' \in {\cal A}} {\exp } \left[ {(1 - \rho ({I_g})){\phi _c}(\alpha ') + \rho ({I_g}){\phi _f}(\alpha ')} \right]}},
\end{equation}
where the denominator makes the average value of $w_\alpha(I_g)$ over $\mathcal{A}$ equal to one, thus preserving the overall response scale while adjusting the relative reliability of different observation scales. Because this weight reflects scale-level reliability, it should affect only the retention priority of candidates rather than their spatial locations or descriptors. Therefore, the keypoint coordinate and descriptor are kept unchanged, and only the detector response is recalibrated as
\begin{equation}
    \widetilde{r}_{i}^{(\alpha )} = {w_\alpha }({I_g})r_i^{(\alpha )}.
\end{equation}
The calibrated score $\widetilde{r}_{i}^{(\alpha)}$ is then used as the priority measure in cross-scale spatial fusion.

\subsubsection{Cross-Scale Feature Fusion}

After response calibration, feature candidates from all scales are collected into a common candidate pool in the original image coordinate frame:
\begin{equation}
    {\cal H}_u = \bigcup\limits_{\alpha  \in {\cal A}} {{\widetilde{\cal H}^{(\alpha )}}}  = \bigcup\limits_{\alpha  \in {\cal A}} {\left\{ {\left( {\widehat {\bf{p}}_i^{(\alpha )},{\bf{d}}_i^{(\alpha )},\widetilde{r}_{i}^{(\alpha )}} \right)} \right\}_{i = 1}^{{N_\alpha }}} ,
\end{equation}
where $\bigcup$ denotes the union of feature candidates over all scales. Since the same acoustic structure or texture element may activate nearby responses at multiple scales, directly retaining all candidates would introduce redundant detections. Therefore, a score-aware spatial fusion operator is used to obtain a compact feature set.

Let $\operatorname{F}_{\tau_f}(\cdot)$ denote the spatial fusion operator with fusion radius $\tau_f$. Given the candidate pool $\mathcal{H}$, the operator first sorts all candidates in descending order according to their calibrated responses. A candidate is retained if it has not been suppressed by any previously retained candidate. Once a candidate is retained, all remaining candidates within the fusion radius are suppressed:
\begin{equation}
    {\left\| {{{\widehat {\bf{p}}}_i} - {{\widehat {\bf{p}}}_j}} \right\|_2} < {\tau _f}.
\end{equation}
The final fused scale-space feature set is then written as
\begin{equation}
    {{\cal H}^*} = {{\mathop{\rm F}\nolimits} _{{\tau _f}}}\left( {\cal H}_u \right) = \left\{ {\left( {{{\bf{p}}_j},{{\bf{d}}_j},\widetilde{r}_{j}} \right)} \right\}_{j = 1}^N.
\end{equation}
where $\mathbf{p}_j$ denotes the selected keypoint coordinate in the original image frame. Each retained feature inherits the descriptor and calibrated response of the selected scale-space candidate. In this way, cross-scale fusion performs content-adaptive candidate selection by prioritizing candidates according to calibrated responses and suppressing redundant nearby detections. The resulting scale-space feature set $\mathcal{H}^{*}$ is then used as the input to the subsequent sparse matching stage.

\subsection{Context-Consistent Correspondence Generation}

\begin{figure*}[htbp]
	\centering	\includegraphics[width=1\linewidth]{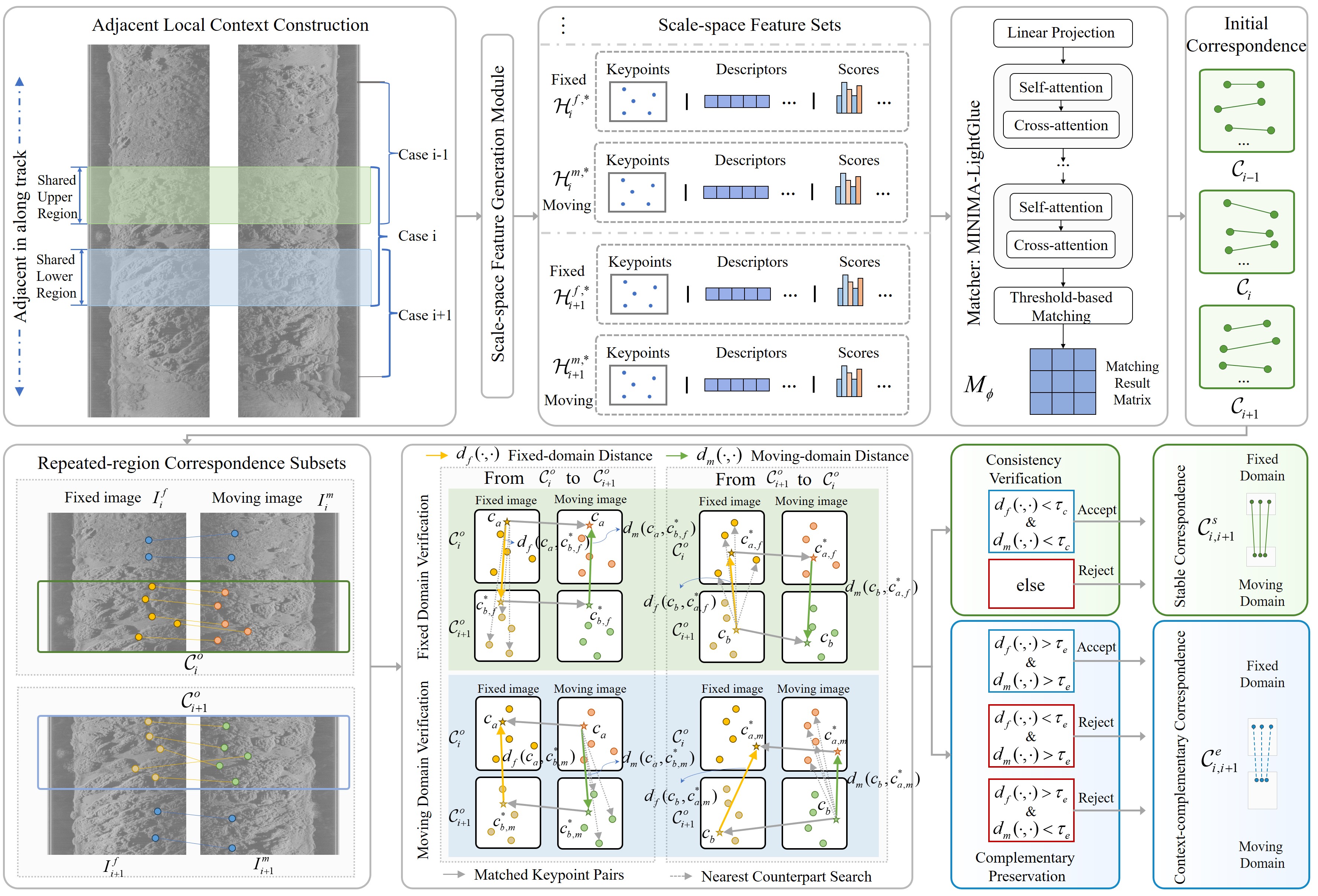}
    \caption{Context-consistent correspondence generation. Adjacent local fixed--moving cases are constructed with shared along-track regions and processed by the scale-space feature generation module to obtain feature sets for each context. A pretrained MINIMA-LightGlue matcher generates initial correspondences, from which repeated-region subsets are extracted and transformed into a common coordinate frame. Cross-context dual-domain verification compares these subsets in both fixed and moving domains to retain reproduced fixed--moving relations as stable correspondences and preserve sufficiently separated non-conflicting relations as context-complementary correspondences. The stable subset $\mathcal{C}_{i,i+1}^{s}$ and context-complementary subset $\mathcal{C}_{i,i+1}^{e}$ are then used for subsequent correspondence consolidation and alignment.}
	\label{method03}
\end{figure*}

Given the fused scale-space feature set $\mathcal{H}^{*}$, correspondences are generated within fixed--moving matching cases, i.e., local SSS observation pairs consisting of a fixed observation as the reference and a moving observation to be aligned, rather than a globally aligned field of view. Each local matching case is treated as a local context because it provides a specific surrounding acoustic observation and feature configuration for correspondence generation. As shown in Fig. \ref{method03}, adjacent local contexts are arranged to share repeated seabed regions in both the fixed and moving domains, so that the same fixed--moving relation can be examined under different local contexts. In these shared repeated regions, correspondences reproduced across adjacent local contexts are treated as stable matches, while non-reproduced correspondences are retained only when they do not conflict with matches from neighboring local contexts. These two types of correspondences provide the refined support for consolidation and alignment.

\subsubsection{Adjacent Local Context Construction}

Let $(I_i^f,I_i^m)$ denote the $i$-th local matching case, and let $(I_{i+1}^f,I_{i+1}^m)$ denote its adjacent local matching case. Adjacent local matching cases are arranged to share an along-track interval in both the fixed and moving domains. With the local observation height $H$ and the shared height $H_o$, the along-track shift between two adjacent local cases is defined as
\begin{equation}
    \Delta y = H-H_o,
\end{equation}
here, a 50\% along-track overlap setting is adopted, i.e., $H_o=H/2$ and $\Delta y=H/2$. Therefore, the lower half of the $i$-th local context and the upper half of the $(i+1)$-th local context correspond to the same along-track interval and form a shared repeated region. Symmetrically, except for boundary cases, the upper part of the $i$-th local case is repeated in the lower part of the $(i-1)$-th local case. Thus, each interior local matching case is associated with repeated regions from both neighboring directions.

For the local matching case $(I_i^f,I_i^m)$, the scale-space feature generation module is applied separately to the fixed and moving observations to produce scale-space feature sets:
\begin{small}
\begin{equation}
    \mathcal{H}_{i}^{f,*}
    =
    \left\{
    \left(
    \mathbf{p}_{i,u}^{f},
    \mathbf{d}_{i,u}^{f},
    \widetilde{r}_{i,u}^{f}
    \right)
    \right\}_{u=1}^{N_i^f},
    \mathcal{H}_{i}^{m,*}
    =
    \left\{
    \left(
    \mathbf{p}_{i,v}^{m},
    \mathbf{d}_{i,v}^{m},
    \widetilde{r}_{i,v}^{m}
    \right)
    \right\}_{v=1}^{N_i^m}.
\end{equation}
\end{small}

The same operation is performed for the adjacent matching case $(I_{i+1}^{f},I_{i+1}^{m})$, producing $\mathcal{H}_{i+1}^{f,*}$ and $\mathcal{H}_{i+1}^{m,*}$; the preceding adjacent case is handled in the same manner.

\subsubsection{Sparse Correspondence Generation}


Given the scale-space feature sets of each matching case, a pretrained MINIMA-LightGlue matcher \cite{A27,A29} is used to generate initial sparse correspondences. Here, MINIMA-LightGlue denotes the pretrained MINIMA matching configuration built on LightGlue-based correspondence reasoning, with all matcher parameters kept fixed during inference. For the $i$-th matching case, this process is formulated as
\begin{equation}
    \mathcal{C}_{i}
    =
    M_{\phi}
    \left(
    \mathcal{H}_{i}^{f,*},
    \mathcal{H}_{i}^{m,*}
    \right)
    =
    \left\{
    \left(
    \mathbf{p}_{i,k}^{f},
    \mathbf{p}_{i,k}^{m},
    s_{i,k}
    \right)
    \right\}_{k=1}^{K_i},
\end{equation}
where $M_{\phi}$ denotes the pretrained matcher with fixed parameters $\phi$ \cite{A27, A29}, $\mathbf{p}_{i,k}^{f}$ and $\mathbf{p}_{i,k}^{m}$ are the matched keypoints in the fixed and moving images, respectively, and $s_{i,k}$ is the matching confidence score. The same matching process is applied to the adjacent matching case, yielding $\mathcal{C}_{i+1}$ and $\mathcal{C}_{i-1}$.

To compare the correspondence results from adjacent contexts, we extract the correspondences located in their repeated regions and represent them in the common coordinate frame of the $i$-th matching case. Here, the adjacent pair $(i,i+1)$ is used for formulation, and the preceding pair $(i-1,i)$ follows the same procedure. The coordinates from the $i$-th case are kept unchanged, whereas those from the adjacent case are shifted by the along-track offset $\boldsymbol{\delta}=(0,\Delta y)^T$:
\begin{equation}
\begin{cases}
    \mathbf{q}_{i,k}^{f} = \mathbf{p}_{i,k}^{f},\\
    \mathbf{q}_{i,k}^{m} = \mathbf{p}_{i,k}^{m},
\end{cases}
\quad
\begin{cases}
    \mathbf{q}_{i+1,k}^{f} = \mathbf{p}_{i+1,k}^{f} + \boldsymbol{\delta},\\
    \mathbf{q}_{i+1,k}^{m} = \mathbf{p}_{i+1,k}^{m} + \boldsymbol{\delta}.
\end{cases}
\end{equation}

The resulting repeated-region correspondence subsets are denoted as
\begin{equation}
    \mathcal{C}_{i}^{o}
    =
    \left\{
    \left(
    \mathbf{q}_{i,k}^{f},
    \mathbf{q}_{i,k}^{m},
    s_{i,k}
    \right)
    \right\},
    \mathcal{C}_{i+1}^{o}
    =
    \left\{
    \left(
    \mathbf{q}_{i+1,k}^{f},
    \mathbf{q}_{i+1,k}^{m},
    s_{i+1,k}
    \right)
    \right\},
\end{equation}
where $\mathbf{q}^{f}$ and $\mathbf{q}^{m}$ are the matched keypoints in the fixed and moving images, respectively, and only correspondences whose fixed and moving keypoints both lie in the shared repeated regions are included.

\subsubsection{Cross-Context Correspondence Verification}

The repeated-region correspondence subsets $\mathcal{C}_{i}^{o}$ and $\mathcal{C}_{i+1}^{o}$ are compared through two criteria. The first criterion verifies whether a fixed--moving relation is reproduced across adjacent contexts, while the second preserves non-reproduced correspondences only when they are separated from neighboring-context matches in both image domains.

For two correspondences from adjacent contexts,
\begin{equation}
    c_a=\left(\mathbf{q}_{a}^{f},\mathbf{q}_{a}^{m},s_a\right)\in\mathcal{C}_{i}^{o}, \quad c_b=\left(\mathbf{q}_{b}^{f},\mathbf{q}_{b}^{m},s_b\right)\in\mathcal{C}_{i+1}^{o},
\end{equation}
their fixed-domain and moving-domain distances are defined as
\begin{equation}
    d_f(c_a,c_b)=\left\|\mathbf{q}_{a}^{f}-\mathbf{q}_{b}^{f}\right\|_2,  d_m(c_a,c_b)=\left\|\mathbf{q}_{a}^{m}-\mathbf{q}_{b}^{m}\right\|_2 .
\end{equation}

To identify reproduced fixed--moving relations, a query correspondence $c_a\in\mathcal{C}_{i}^{o}$ is compared with the adjacent-context subset $\mathcal{C}_{i+1}^{o}$. Its nearest counterpart is first searched according to the fixed-domain distance:
\begin{equation}
    c_{b,f}^{*}=\arg\min_{c_b\in\mathcal{C}_{i+1}^{o}} d_f(c_a,c_b),
\end{equation}
then, the fixed-guided verification indicator is defined as
\begin{equation}
    \Gamma_f\left(c_a;\mathcal{C}_{i+1}^{o}\right)=\mathbb{I}\left[d_f(c_a,c_{b,f}^{*})<\tau_c \ \land\ d_m(c_a,c_{b,f}^{*})<\tau_c\right],
\end{equation}
where $\tau_c$ is the cross-context consistency threshold and $\mathbb{I}[\cdot]$ denotes the indicator function. This verification accepts $c_a$ only when the correspondence nearest in the fixed domain is also spatially consistent in the moving domain.

Similarly, using the moving domain as the query domain, the nearest counterpart is identified by
\begin{equation}
    c_{b,m}^{*}=\arg\min_{c_b\in\mathcal{C}_{i+1}^{o}} d_m(c_a,c_b),
\end{equation}
and the moving-guided verification indicator is defined as
\begin{equation}
    \Gamma_m\left(c_a;\mathcal{C}_{i+1}^{o}\right)=\mathbb{I}\left[d_m(c_a,c_{b,m}^{*})<\tau_c \ \land\ d_f(c_a,c_{b,m}^{*})<\tau_c\right].
\end{equation}

The stable subset contributed by $\mathcal{C}_{i}^{o}$ with respect to $\mathcal{C}_{i+1}^{o}$ is therefore defined as
\begin{equation}
    \mathcal{S}_{i\rightarrow i+1}=\left\{c_a\in\mathcal{C}_{i}^{o}\mid \Gamma_f\left(c_a;\mathcal{C}_{i+1}^{o}\right)=1 \ \lor\ \Gamma_m\left(c_a;\mathcal{C}_{i+1}^{o}\right)=1\right\}.
\end{equation}
The same verification is performed in the reverse context direction, where correspondences from $\mathcal{C}_{i+1}^{o}$ are verified against $\mathcal{C}_{i}^{o}$:
\begin{equation}
    \mathcal{S}_{i+1\rightarrow i}=\left\{c_b\in\mathcal{C}_{i+1}^{o}\mid \Gamma_f\left(c_b;\mathcal{C}_{i}^{o}\right)=1 \ \lor\ \Gamma_m\left(c_b;\mathcal{C}_{i}^{o}\right)=1\right\}.
\end{equation}

Thus, the final stable correspondence subset is obtained as
\begin{equation}
    \mathcal{C}_{i,i+1}^{s}=\operatorname{CC}_{\tau_c}\left(\mathcal{C}_{i}^{o},\mathcal{C}_{i+1}^{o}\right)=\mathcal{S}_{i\rightarrow i+1}\cup\mathcal{S}_{i+1\rightarrow i},
\end{equation}
where $\operatorname{CC}_{\tau_c}(\cdot)$ denotes the cross-context dual-domain consistency operation. It retains correspondences that are reproduced as consistent fixed--moving relations across adjacent contexts.

For non-reproduced correspondence preservation, a candidate correspondence is retained only if it is sufficiently separated from the adjacent-context subset in both image domains, so that it can provide complementary spatial support without conflicting with neighboring-context matches. For $c_a\in\mathcal{C}_{i}^{o}$, the nearest fixed-domain and moving-domain distances to $\mathcal{C}_{i+1}^{o}$ are computed as
\begin{equation}
    \begin{array}{l}
    d_f^{\min }\left( {{c_a};{\cal C}_{i + 1}^o} \right) = {\min _{{c_b} \in {\cal C}_{i + 1}^o}}{d_f}({c_a},{c_b}),\\
    d_m^{\min }\left( {{c_a};{\cal C}_{i + 1}^o} \right) = {\min _{{c_b} \in {\cal C}_{i + 1}^o}}{d_m}({c_a},{c_b}),
    \end{array}
\end{equation}

A non-conflicting candidate is identified using a dual-domain separation criterion:
\begin{equation}
    \Psi\left(c_a;\mathcal{C}_{i+1}^{o}\right)=\mathbb{I}\left[d_f^{\min}\left(c_a;\mathcal{C}_{i+1}^{o}\right)>\tau_e \ \land\ d_m^{\min}\left(c_a;\mathcal{C}_{i+1}^{o}\right)>\tau_e\right],
\end{equation}
where $\tau_e$ is the separation threshold. This criterion retains a non-reproduced correspondence only when it is sufficiently separated from the adjacent-context subset in both the fixed and moving domains. If a candidate is close to the adjacent-context subset in only one domain, it is regarded as a one-sided conflict and is rejected, because such a case may indicate a shifted or ambiguous match rather than an independent complementary correspondence.

Accordingly, the context-complementary subset contributed by $\mathcal{C}_{i}^{o}$ with respect to $\mathcal{C}_{i+1}^{o}$ is defined as
\begin{equation}
    \mathcal{E}_{i\rightarrow i+1}=\left\{c_a\in\mathcal{C}_{i}^{o}\mid \Psi\left(c_a;\mathcal{C}_{i+1}^{o}\right)=1\right\}.
\end{equation}
The same preservation rule is applied in the reverse context direction:
\begin{equation}
    \mathcal{E}_{i+1\rightarrow i}=\left\{c_b\in\mathcal{C}_{i+1}^{o}\mid \Psi\left(c_b;\mathcal{C}_{i}^{o}\right)=1\right\}.
\end{equation}

The final context-complementary correspondence subset is obtained by combining the two directional contributions:
\begin{equation}
    \mathcal{C}_{i,i+1}^{e}=\operatorname{CE}_{\tau_e}\left(\mathcal{C}_{i}^{o},\mathcal{C}_{i+1}^{o}\right)=\mathcal{E}_{i\rightarrow i+1}\cup\mathcal{E}_{i+1\rightarrow i},
\end{equation}
where $\operatorname{CE}_{\tau_e}(\cdot)$ denotes the bidirectional context complementary preservation operation.

Together, $\mathcal{C}_{i,i+1}^{s}$ and $\mathcal{C}_{i,i+1}^{e}$ form the refined correspondence set for the repeated region between this pair of adjacent contexts. The stable subset $\mathcal{C}_{i,i+1}^{s}$ provides reproduced fixed--moving relations, while the context-complementary subset $\mathcal{C}_{i,i+1}^{e}$ adds non-reproduced but non-conflicting spatial support. Both subsets are retained for subsequent correspondence consolidation and confidence-guided filtering.

\subsection{Correspondence Consolidation and Filtering}

After cross-context correspondence verification, each adjacent repeated region provides two refined correspondence subsets: the stable subset $\mathcal{C}_{i,i+1}^{s}$ and the context-complementary subset $\mathcal{C}_{i,i+1}^{e}$. For the $i$-th matching case, refined correspondences are obtained from both neighboring directions, namely the repeated region shared with the previous adjacent context and that shared with the next adjacent context. The full refined set is therefore consolidated as
\begin{equation}
    \mathcal{C}_{i}^{r}
    =
    \left(
    \mathcal{C}_{i-1,i}^{s}
    \cup
    \mathcal{C}_{i-1,i}^{e}
    \right)
    \cup
    \left(
    \mathcal{C}_{i,i+1}^{s}
    \cup
    \mathcal{C}_{i,i+1}^{e}
    \right),
\end{equation}
where $\mathcal{C}_{i}^{r}$ denotes the consolidated correspondence set for the $i$-th matching case. Since the same fixed--moving correspondence may be introduced from different adjacent-context verifications, repeated entries with identical rounded coordinate pairs are further removed after consolidation, yielding the duplicate-suppressed set $\bar{\mathcal{C}}_{i}^{r}$.

To further remove low-confidence candidates, confidence-guided filtering is performed according to the score distribution of each matching case. Instead of using a fixed global score threshold, a local threshold is computed from the matching scores in $\bar{\mathcal{C}}_{i}^{r}$:
\begin{equation}
    T_i = Q_q(\mathcal{S}_i),
    \quad
    \mathcal{S}_i =
    \left\{
    s_{i,k}
    \mid
    c_{i,k} \in \bar{\mathcal{C}}_{i}^{r}
    \right\},
\end{equation}
where $Q_q(\cdot)$ denotes the $q$-th quantile operator, and $s_{i,k}$ is the matching confidence score of correspondence $c_{i,k}$. The final correspondence set is obtained by retaining correspondences whose confidence scores are not lower than the local threshold:
\begin{equation}
    \mathcal{C}_{i}^{*}
    =
    \left\{
    c_{i,k}\in\bar{\mathcal{C}}_{i}^{r}
    \mid
    s_{i,k}\geq T_i
    \right\},
\end{equation}
where $\mathcal{C}_{i}^{*}$ denotes the final refined correspondence set for the $i$-th matching case. This set is then used for RANSAC-based geometric verification and alignment.

\subsection{RANSAC-based Alignment}

Given the final refined correspondence set $\mathcal{C}_{i}^{*}$, the geometric relation of the $i$-th matching case is modeled by a homography. For each correspondence
$c_{i,k}=(\mathbf{q}_{i,k}^{f},\mathbf{q}_{i,k}^{m},s_{i,k})\in\mathcal{C}_{i}^{*}$, the moving-to-fixed relation is written as
\begin{equation}
    \lambda
    \tilde{\mathbf{q}}_{i,k}^{f}
    =
    \mathbf{H}_{i}
    \tilde{\mathbf{q}}_{i,k}^{m},
\end{equation}
where
$\tilde{\mathbf{q}}_{i,k}^{f}=[x_{i,k}^{f},y_{i,k}^{f},1]^T$ and
$\tilde{\mathbf{q}}_{i,k}^{m}=[x_{i,k}^{m},y_{i,k}^{m},1]^T$
are the homogeneous coordinates of the matched fixed and moving points, respectively. $\mathbf{H}_{i}$ denotes the homography from the moving observation frame to the fixed observation frame, and $\lambda$ is the homogeneous scale factor.

RANSAC \cite{A30} is then used to estimate a geometrically consistent homography from $\mathcal{C}_{i}^{*}$. For a candidate homography $\mathbf{H}$, the reprojection error of $c_{i,k}$ is defined as
\begin{equation}
    e_{i,k}(\mathbf{H})
    = \left|\pi\left(\mathbf{H}\tilde{\mathbf{q}}_{i,k}^{m}\right)-\mathbf{q}_{i,k}^{f}\right|_2,
\end{equation}
where $\pi(\cdot)$ denotes homogeneous coordinate normalization. The optimal homography is selected by maximizing the number of inlier correspondences:
\begin{equation}
    \mathbf{H}_{i}^{*}
    =\arg\max_{\mathbf{H}}
    \left|\left\{ c_{i,k}\in\mathcal{C}_{i}^{*}
    \mid e_{i,k}(\mathbf{H})<\tau_r
    \right\}
    \right|,
\end{equation}
where $\tau_r$ is the reprojection threshold. With this moving-to-fixed convention, $\mathbf{H}_{i}^{*}$ aligns the moving observation to the fixed observation frame. For image generation, backward resampling is used to generate the warped moving observation in the fixed frame:
\begin{equation}
    \hat{I}_{i}^{m}(\mathbf{x})=I_{i}^{m}\left(\pi\left(\left(\mathbf{H}_{i}^{*}\right)^{-1}\tilde{\mathbf{x}}\right)\right),\quad\mathbf{x}\in\Omega_{i}^{f},
\end{equation}
where $\tilde{\mathbf{x}}=[x,y,1]^T$ corresponds to the homogeneous coordinate of a pixel in the fixed frame, $\Omega_{i}^{f}$ denotes the fixed local image domain, and $\hat{I}_{i}^{m}$ is the moving observation warped into the fixed frame. Estimating $\mathbf{H}_{i}^{*}$ independently for each fixed--moving matching case allows the alignment to adapt to spatially varying local geometric relations. 

With the complete SC-Match pipeline defined, the following section evaluates its performance on two SSS datasets acquired under different survey conditions and with different sonar systems and acquisition platforms.

\section{Experiments}

\subsection{Data Description}


Two SSS datasets acquired from different survey sectors and sonar platforms are used in this work, as summarized in Table \ref{datasets}. Dataset I was collected by the Girona1000 AUV equipped with a Marine Sonic Arc Scout MK II SSS. The geo-referencing of this dataset is supported by USBL-assisted positioning and attitude/heading measurements from an OCTANS fiber-optic gyroscope, which enable the construction of a dense fixed-to-moving reference field for each evaluated matching case. This reference field provides reference correspondences for matched-point evaluation and supports the generation of reference warped observations for image-level alignment assessment. Dataset II was acquired from a survey-vessel platform equipped with a Klein 3000H SSS. Although USBL-based navigation is available for Dataset II, its accuracy is insufficient for dense reference-field construction and therefore cannot be regarded as ground truth. Dataset II is used for qualitative generalization evaluation, where the USBL-based result serves only as a coarse visual reference for assessing whether the proposed method produces geometrically reasonable alignment under unseen acquisition conditions.

\begin{table}[t]
    \centering
    \caption{Measurement systems and acquisition conditions of the two side-scan sonar datasets.}
    \label{datasets}
    \renewcommand{\arraystretch}{1.2}
    \setlength{\tabcolsep}{1pt}

    \begin{tabular}{lcc}
    \toprule
    \textbf{Item} & \textbf{Dataset I} & \textbf{Dataset II} \\
    \midrule
    Survey Sector & Sector N08~\cite{A31} & Sector N07~\cite{A31} \\
    Sonar Model & Marine Sonic Arc Scout MK II & Klein 3000H \\
    Operating Frequency & 900 kHz & 500 kHz \\
    Slant Range & 30--80 m & 50--100 m \\
    Sonar Altitude & $\sim$10\% of range & $\sim$10\% of range \\
    Navigation Reliability & High & Limited \\
    Dense Reference Field & Available & Unavailable \\
    Evaluation Role & Quantitative evaluation & Qualitative evaluation \\
\bottomrule
\end{tabular}
\end{table}

\begin{figure*}[htbp]
	\centering	\includegraphics[width=1\linewidth]{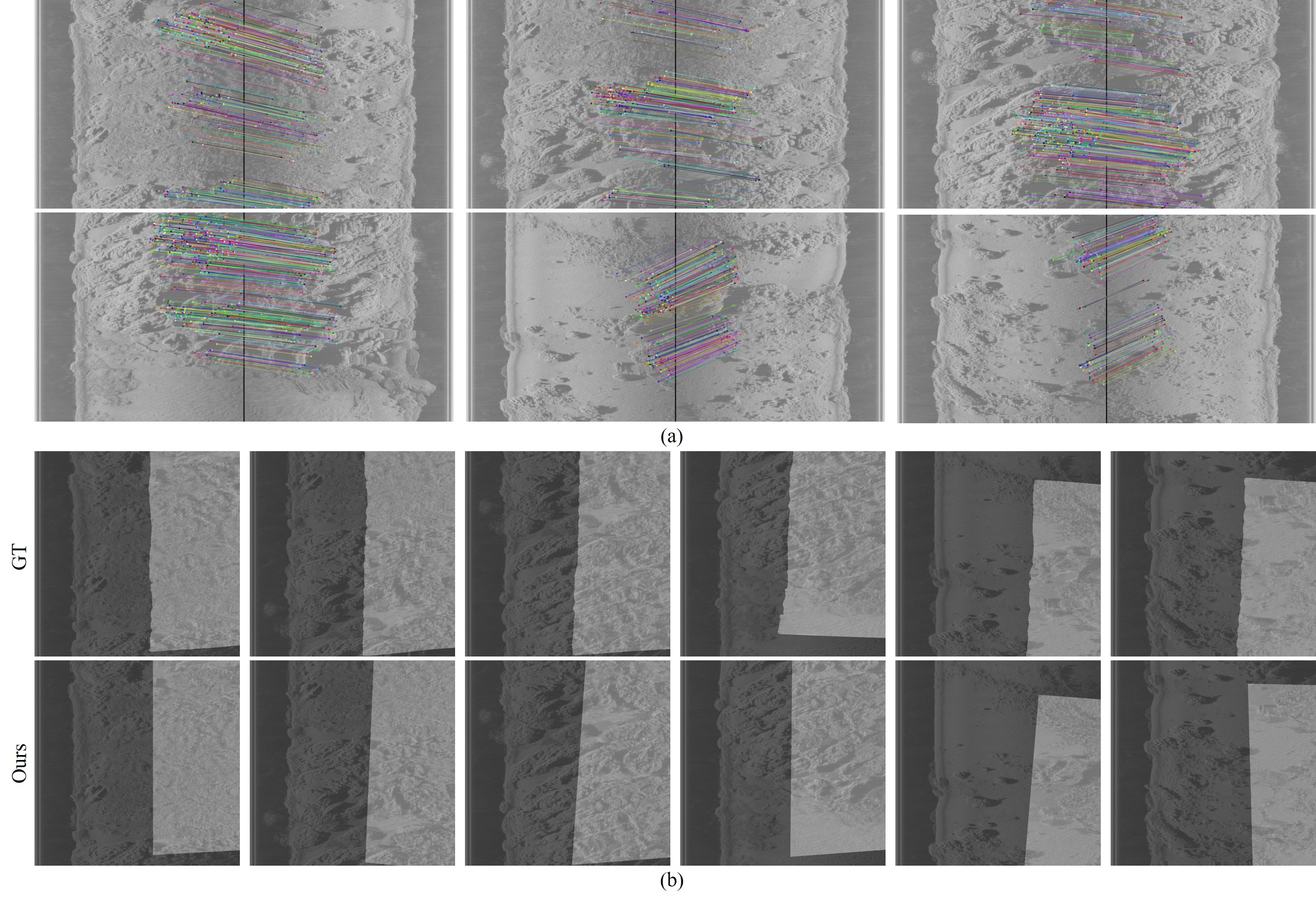}
    \caption{Visualization results of the proposed method. (a) Final refined correspondence sets. The correspondences are mainly concentrated in valid overlapping regions and show coherent displacement directions within each matching case. (b) Image-level warping comparison. The first row shows the reference warped moving observations, and the second row shows the warped results estimated from the proposed correspondences. The proposed results exhibit similar warped footprints, covered regions, and global orientations to the reference warps, with minor differences in locally distorted regions.}
	\label{fig01}
\end{figure*}	

\begin{figure*}[htbp]
	\centering	\includegraphics[width=1\linewidth]{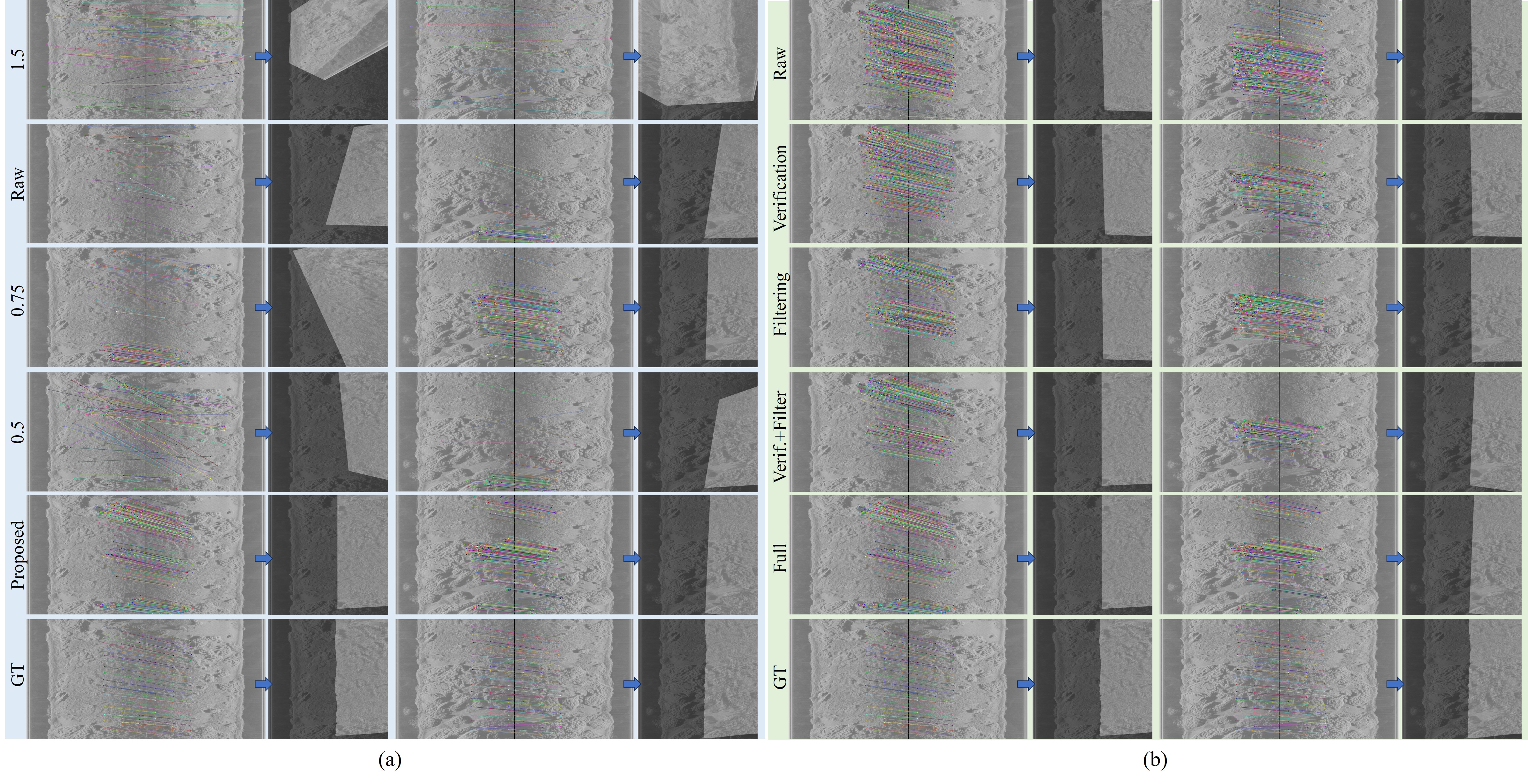}
    \caption{Visualization results of the ablation study. (a) Ablation on scale-space feature generation. The proposed setting produces correspondence distributions and warped footprints closer to the sampled reference correspondences and reference warps (GT) than the single-scale variants. (b) Ablation on correspondence generation and refinement. The full setting provides the most coherent correspondence directions and warped footprints closest to the reference results (GT).}
	\label{fig02}
\end{figure*}	

\subsection{Evaluation Metrics}

The evaluation metrics are organized into two groups. The first group evaluates matched correspondences using the dense fixed-to-moving reference field constructed from Dataset I. The second group evaluates image-level alignment quality by comparing the estimated warped moving observation with the reference warped observation.

\subsubsection{Reference-Field-Based Matched-Point Metrics}

Let
\begin{equation}
    \mathcal{C}
    =\left\{ c_k =\left(\mathbf{q}_{k}^{f},\mathbf{q}_{k}^{m}
    \right)\right\}_{k=1}^{N}
\end{equation}
denote the correspondence set produced by an evaluated matching method, where $\mathbf{q}_{k}^{f}$ and $\mathbf{q}_{k}^{m}$ are the matched points in the fixed and moving observation frames, respectively. Let $\mathcal{G}(\cdot)$ denote the dense fixed-to-moving reference field. The pixel error of the $k$-th matched correspondence is defined as
\begin{equation}
    e_{k}^{\mathrm{px}}
    =\left\| \mathbf{q}_{k}^{m}
    -\mathcal{G}\left(\mathbf{q}_{k}^{f} \right)\right\|_{2}.
\end{equation}

The evaluated matched-point metrics include:

\begin{itemize}
    \item Mean matching error. The mean matching error measures the average deviation between the estimated matched correspondences and the reference-field-derived correspondences:
    \begin{equation}
        \bar{e}_{\mathrm{px}}=\frac{1}{N}\sum_{k=1}^{N}e_{k}^{\mathrm{px}},
    \end{equation}
    in addition, the corresponding metric-space mean error is reported as $\bar{e}_{\mathrm{m}}$.

    \item Standard deviation of matching error. The standard deviation measures the dispersion of the matched-point errors:
    \begin{equation}
        \sigma_{e}=\sqrt{\frac{1}{N-1}\sum_{k=1}^{N}\left(e_{k}^{\mathrm{px}}-\bar{e}_{\mathrm{px}}\right)^{2}} .
    \end{equation}

    \item Correct matching ratio. The correct matching ratio measures the proportion of matched correspondences whose pixel errors are below a predefined tolerance. In this work, the tolerance is set to $\tau_{\mathrm{px}}=30$ pixels (corresponding to $0.45$ m):
    \begin{equation}
        R_{\mathrm{c}} =\frac{1}{N}\sum_{k=1}^{N}\mathbb{I}\left(e_{k}^{\mathrm{px}} <\tau_{\mathrm{px}}\right),
    \end{equation}
    where $\mathbb{I}(\cdot)$ denotes the indicator function. Lower $\bar{e}_{\mathrm{px}}$, $\bar{e}_{\mathrm{m}}$, and $\sigma_{e}$ indicate more accurate and stable correspondences, while a higher $R_{\mathrm{c}}$ indicates a larger proportion of reliable matches. 
\end{itemize}

\subsubsection{Reference-Warp-based Image-level Metrics}

The image-level alignment quality is evaluated by comparing the estimated warped moving observation $\hat{I}_{i}^{m}$ with the reference warped moving observation $I_{i,\mathrm{ref}}^{m}$ in their common valid region $\Omega_{i}$. The image-level metrics include:

\begin{itemize}
    \item Root mean square error. RMSE measures the pixel-wise intensity discrepancy between the estimated and reference warped observations:
    \begin{equation}
        \mathrm{RMSE}=\sqrt{\frac{1}{|\Omega_{i}|}\sum_{\mathbf{x}\in\Omega_{i}}\left(\hat{I}_{i}^{m}(\mathbf{x}) -I_{i,\mathrm{ref}}^{m}(\mathbf{x})\right)^{2}} .
    \end{equation}

    \item Mutual information (MI). MI measures the statistical dependence between the estimated and reference warped observations:
    \begin{equation}
        \mathrm{MI}=\sum_{a}\sum_{b} p_{\hat{I},I_{\mathrm{ref}}}(a,b)\log\frac{p_{\hat{I},I_{\mathrm{ref}}}(a,b)}{ p_{\hat{I}}(a) p_{I_{\mathrm{ref}}}(b) } .
    \end{equation}
    here, $p_{\hat{I},I_{\mathrm{ref}}}(a,b)$ denotes the joint intensity distribution of $\hat{I}_{i}^{m}$ and $I_{i,\mathrm{ref}}^{m}$, while $p_{\hat{I}}(a)$ and $p_{I_{\mathrm{ref}}}(b)$ denote their marginal distributions.

    \item Learned perceptual image patch similarity (LPIPS). LPIPS \cite{A32} measures the deep-feature discrepancy between the estimated and reference warped observations:
    
    \begin{footnotesize}
    \begin{equation}
        \mathrm{LPIPS}=\sum_{l}\frac{1}{H_{l}W_{l}}\sum_{\mathbf{x}}\left\|\mathbf{w}_{l} \odot \left( \phi_{l} \left(  \hat{I}_{i}^{m}  \right) (\mathbf{x})   -   \phi_{l}  \left(   I_{i,\mathrm{ref}}^{m}  \right) (\mathbf{x}) \right) \right\|_{2}^{2}.
    \end{equation}
    \end{footnotesize}here, $\phi_{l}(\cdot)$ denotes the feature map extracted from the $l$-th layer of the LPIPS network, $\mathbf{w}_{l}$ is the learned channel-wise weighting vector, and $H_{l}$ and $W_{l}$ are the spatial dimensions of the feature map. Lower $\mathrm{RMSE}$ and $\mathrm{LPIPS}$ and higher $\mathrm{MI}$ indicate better agreement with the reference warped moving observation.
    
\end{itemize}

\subsection{Experimental Settings}

The candidate observation-scale set is $\mathcal{A}=\{0.5,0.75,1.0,1.5,2.0\}$. The frozen SuperPoint extractor is used with an NMS radius of $4$ and a detection threshold of $0.01$, and the cross-scale fusion radius is set to $\tau_f=2.0$ pixels. For correspondence verification, the consistency threshold and the context-complementary separation threshold are set to $\tau_c=20$ pixels and $\tau_e=80$ pixels, respectively. After correspondence consolidation and duplicate suppression, confidence-guided filtering is performed with $q=0.5$. A homography is estimated for each matching case using RANSAC with a reprojection threshold of $\tau_r=20$ pixels. 

\section{Results}

\subsection{Visualization Results}

Fig. \ref{fig01} shows representative visualization results of the proposed method. Fig. \ref{fig01}(a) presents the final refined correspondence sets. The correspondences are mainly located in valid overlapping regions and exhibit coherent displacement directions within each matching case. Rather than being uniformly distributed over the entire image, they are concentrated around seabed structures, textured terrain, and acoustic intensity variations, where reliable local cues are more likely to be generated. This distribution is consistent with the proposed scale-space feature generation and context-consistent correspondence generation, where content-adaptive response calibration improves feature selection and cross-context correspondence verification suppresses unstable fixed--moving relations across adjacent local contexts.

Fig. \ref{fig01}(b) compares the reference warped moving observations with the warped results estimated from the proposed correspondences. Across the six representative cases, the proposed results show similar warped footprints, covered regions, and global orientations to the reference warps. This visual consistency indicates that the final refined correspondences provide effective geometric support for RANSAC-based alignment. Minor differences remain near warped boundaries and locally distorted regions, mainly due to residual non-rigid distortions that cannot be fully represented by a homography. Overall, the proposed method recovers geometrically reasonable alignment results close to the reference warps.

\begin{table*}[t]
    \centering
    \caption{Ablation study on the proposed scale-space feature generation and correspondence-side components. The correspondence errors are reported in pixels and meters. $R_c$ is computed using the 30 px error threshold.}
    \label{ablation}
    \renewcommand{\arraystretch}{1.2}
    \setlength{\tabcolsep}{10.7pt}
    \begin{tabular}{llccccccc}
    \toprule
    \textbf{Module} & \textbf{Setting} & $\bar{e}_{\mathrm{px}}\downarrow$ & $\bar{e}_{\mathrm{m}}\downarrow$ & $\sigma_e\downarrow$ & $R_c(\%)\uparrow$ & RMSE$\downarrow$ & MI$\uparrow$ & LPIPS$\downarrow$ \\
    \midrule
    \multirow{5}{*}{Feature generation}
    & $1.5$-scale & 977.83 & 14.67 & 257.06 & 0.00 & 36.12 & 0.694 & 0.219 \\
    & Raw-scale & 837.76 & 12.57 & 133.74 & 0.00 & 23.51 & 0.663 & 0.205 \\
    & $0.75$-scale & 42.64 & 0.64 & 14.11 & 49.71 & 22.24 & 0.693 & 0.165 \\
    & $0.5$-scale & 267.45 & 4.01 & 244.00 & 18.19 & 33.53 & 0.654 & 0.222 \\
    & Proposed & \textbf{38.57} & \textbf{0.58} & \textbf{9.39} & \textbf{58.29} & \textbf{21.70} & \textbf{0.705} & \textbf{0.164} \\
    \midrule
    \multirow{5}{*}{Correspondence generation}
    & Raw & 46.93 & 0.70 & 21.71 & 38.29 & 22.07 & 0.693 & 0.168 \\
    & Verification & 42.14 & 0.63 & 14.30 & 47.51 & 21.99 & 0.698 & 0.165 \\
    & Filtering & 43.01 & 0.65 & 15.10 & 48.61 & 22.00 & 0.695 & 0.166 \\
    & Verification + filtering & 41.80 & 0.63 & 12.44 & 49.54 & 21.81 & 0.699 & 0.165 \\
    & Full setting & \textbf{38.57} & \textbf{0.58} & \textbf{9.39} & \textbf{58.29} & \textbf{21.70} & \textbf{0.705} & \textbf{0.164} \\
    \bottomrule
    \end{tabular}
\end{table*}

\begin{table*}[t]
    \centering
    \caption{Quantitative comparison with representative pretrained image matching methods. The correspondence errors are reported in pixels. The post-RANSAC errors are denoted by $\bar{e}_{\mathrm{R}}$ and $\sigma_{e,\mathrm{R}}$.}
    \label{sota}
    \renewcommand{\arraystretch}{1.2}
    \setlength{\tabcolsep}{6.8pt}
    \begin{tabular}{llccccccccc}
    \toprule
    \textbf{Category} & \textbf{Method} & $\bar{e}_{\mathrm{px}}\downarrow$ & $\sigma_{e}\downarrow$ & $\bar{e}_{\mathrm{R}}\downarrow$ & $\sigma_{e,\mathrm{R}}\downarrow$ & RMSE$\downarrow$ & MI$\uparrow$ & LPIPS$\downarrow$ & Params (M)$\downarrow$ & Time (ms)$\downarrow$ \\
    \midrule
    \multirow{4}{*}{CNN-based}
    & RIPE \cite{A24}   & 786.01  & 398.42 & 578.28  & 282.90 & 27.00 & 0.692 & 0.354 & 14.81 & 133.64 \\
    & LiftFeat \cite{A33} & 786.10  & 348.55 & 738.22  & 176.43 & 35.24 & 0.669 & 0.320 & 2.00  & \textbf{42.71}  \\
    & XFeat \cite{A23}  & 690.81  & 396.41 & 604.85  & 338.48 & 30.04 & 0.671 & 0.297 & \textbf{1.54}  & 43.74  \\
    & ALIKED \cite{A34} & 461.86  & 264.89 & 406.42  & 315.10 & 35.53 & 0.636 & 0.229 & 12.56 & 61.81  \\
    \midrule
    \multirow{4}{*}{Transformer-based}
    & RDD  \cite{A26}   & 720.21  & 433.93 & 417.66  & 273.64 & 28.09 & 0.673 & 0.304 & 32.90 & 175.24 \\
    & MINIMA \cite{A27} & 1081.66 & 204.40 & 1055.17 & 19.25  & 35.94 & 0.690 & 0.222 & 11.56 & 90.16  \\
    & XoFTR \cite{A35}  & 425.15  & 296.73 & 48.41   & 21.41  & 22.17 & 0.698 & 0.173 & 11.09 & 145.03 \\
    & E-LoFTR \cite{A25} & 249.98  & 237.10 & 59.83   & 37.49  & 21.92 & 0.704 & 0.180 & 15.05 & 68.26  \\
    \midrule
    Proposed
    & SC-Match (Ours)    & \textbf{38.57} & \textbf{9.39} & \textbf{37.87} & \textbf{8.86}
    & \textbf{21.70} & \textbf{0.705} & \textbf{0.164} & 13.15 & 90.79 \\
    \bottomrule
    \end{tabular}
\end{table*}

\begin{figure*}[htbp]
	\centering	\includegraphics[width=1\linewidth]{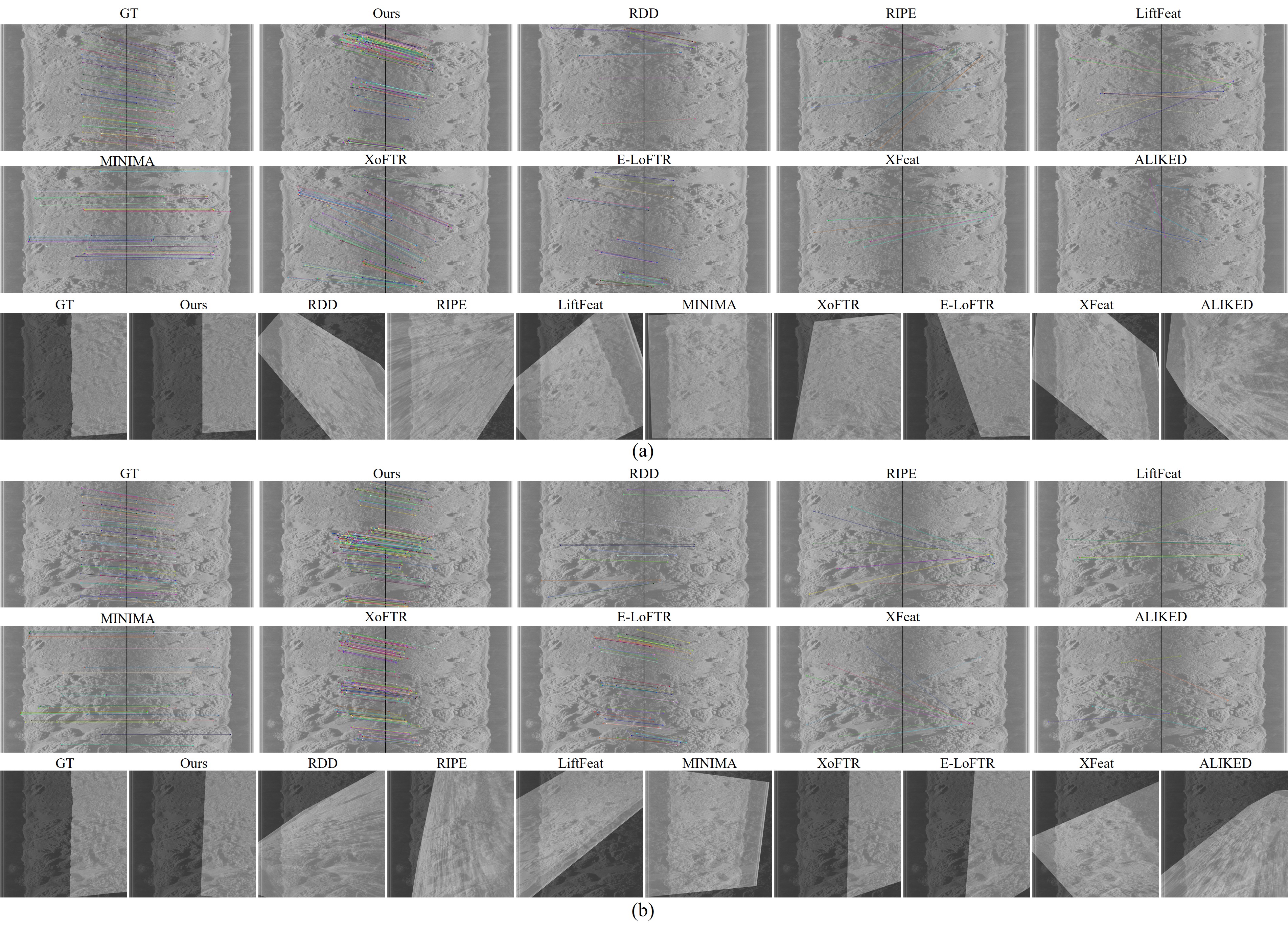}
    \caption{Visual comparison with representative pretrained image matching methods on two SSS mapping cases. Case (a) illustrates matching with relatively sparse local acoustic cues, whereas Case (b) highlights observations dominated by pronounced acoustic shadows. For each case, the first two rows show the correspondence visualizations of the compared methods and the sampled reference correspondences, where the latter indicate the reference global correspondence tendency. The third row shows the corresponding warped moving observations and the reference warp.}
	\label{fig03}
\end{figure*}	

\begin{figure*}[htbp]
	\centering	\includegraphics[width=1\linewidth]{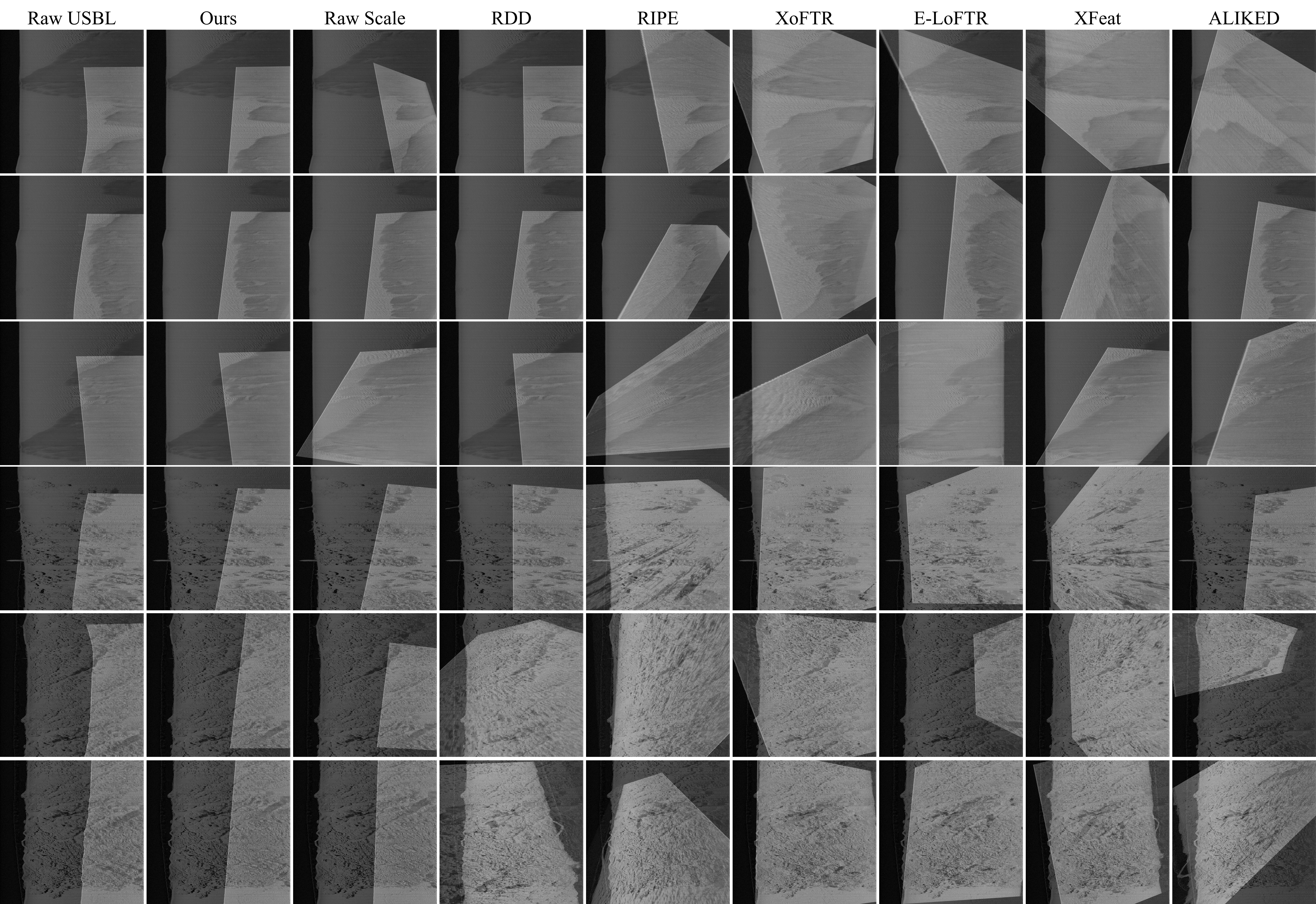}
     \caption{Qualitative cross-platform comparison on Dataset II. Since dense reference fields are unavailable, the Raw USBL-based result is shown only as a coarse reference for the approximate coverage tendency rather than as an exact ground truth. The proposed method produces more regular warped footprints and more stable covered regions than most competing methods under unseen acquisition conditions.}
	\label{fig04}
\end{figure*}	

\subsection{Ablation Study}

Ablation studies are conducted on Dataset I to evaluate the contributions of the scale-space feature generation module and the correspondence-side design.

\subsubsection{Ablation on Feature Generation}

This ablation evaluates the contribution of the proposed scale-space feature generation module. The correspondence generation, consolidation, filtering, and alignment stages are kept unchanged, while only the feature generation strategy is varied. Specifically, the proposed setting is compared with four single-scale variants, denoted as $1.5$ scale, raw scale, $0.75$ scale, and $0.5$ scale. Each single-scale variant extracts features from only one observation scale and therefore removes content-adaptive response calibration and cross-scale feature fusion. In Fig. \ref{fig02}(a), the reference correspondences are sparsely sampled from the dense field to indicate the global correspondence tendency.

As reported in Table \ref{ablation}, the proposed setting achieves the best overall performance in both correspondence accuracy and image-level alignment. Among the single-scale variants, the 0.75-scale setting is the closest competitor but still shows larger error dispersion and weaker image-level consistency, while the other single-scale settings produce less reliable correspondences or alignment. These results indicate that a single observation scale is insufficient to consistently preserve complementary structural and textural cues. Consistent with Fig. \ref{fig02}(a), the quantitative results support the effectiveness of the overall scale-space feature generation design in providing more reliable correspondence support.

\subsubsection{Ablation on Correspondence Generation and Refinement}

This ablation evaluates the contribution of the correspondence-side components. The proposed scale-space feature generation module is fixed, and five settings are compared: raw correspondence generation, cross-context correspondence verification only, confidence-guided filtering only, cross-context correspondence verification with confidence-guided filtering, and the full proposed setting. 

As shown in Table \ref{ablation}, cross-context correspondence verification and confidence-guided filtering both improve the raw correspondence results, and their combination further reduces the correspondence error. The full setting achieves the best overall performance, reducing $\bar{e}_{\mathrm{px}}$ to $38.57$ px and $\sigma_e$ to $9.39$ px, which indicates that stable correspondences and context-complementary correspondences provide complementary support. Specifically, the former preserve reproduced relations across adjacent contexts, while the latter retain non-reproduced but non-conflicting matches. Fig. \ref{fig02}(b) further shows that the full setting produces more coherent correspondence directions and warped footprints closer to the reference, confirming the effectiveness of the proposed refinement design.

\subsection{Comparison with Representative Methods}

SC-Match is compared with representative pretrained image matching methods without sonar-specific fine-tuning. The CNN-based methods include RIPE, LiftFeat, XFeat, and ALIKED, and the transformer-based methods include RDD, MINIMA, XoFTR, and E-LoFTR. For brevity, MINIMA refers to the MINIMA-LoFTR configuration in the tables and figures. For each method, the generated correspondences are evaluated using the dense reference field, and the warped moving observations are evaluated against the reference warps. In addition to the initial correspondence errors, the post-RANSAC errors are also reported to assess the geometric consistency of the correspondences used for alignment. Model parameters and inference time are reported as computational references.

Table \ref{sota} summarizes the quantitative comparison. The proposed method achieves the best overall performance in both correspondence accuracy and image-level alignment. Although XoFTR and E-LoFTR obtain relatively competitive post-RANSAC errors, their much larger initial errors indicate a stronger dependence on robust estimation, whereas SC-Match provides more reliable correspondences before geometric verification. Lightweight methods have lower computational cost but show less favorable accuracy and alignment quality, reflecting the accuracy-oriented trade-off of the proposed method. Fig. \ref{fig03} presents visual comparisons on two representative cases, where the sampled reference correspondences (GT) indicate the reference correspondence tendency derived from the dense reference field. The proposed method produces correspondences that are more concentrated in valid overlapping regions and exhibit more coherent displacement directions, leading to warped footprints closer to the reference. In comparison, several baselines generate sparse, scattered, or directionally inconsistent matches, causing noticeable rotation or displacement deviations in the warped results. These visual results further support the robustness of SC-Match in SSS mapping.

\subsection{Cross-Instrument and Cross-Platform Evaluation}

The generalization capability of the proposed method is further evaluated on Dataset II. Since dense reference fields are unavailable for this dataset, the evaluation is conducted through visual comparison of the warped moving observations. All methods are directly applied without sonar-specific fine-tuning, using the same pretrained models and parameter settings as in Dataset I. As shown in Fig.~\ref{fig04}, the comparison includes the Raw USBL-based result, the proposed method, the raw-scale feature variant, and representative pretrained matching methods. The Raw USBL-based result is used only as a coarse reference for the approximate coverage tendency rather than as an exact ground truth.

Fig.~\ref{fig04} demonstrates visually stable alignment behavior of the proposed method under unseen acquisition conditions. In the first several cases, where the overlap boundaries are visually clear, the Raw USBL-based results provide a generally reasonable global orientation but contain visible artifacts and footprint offsets, reflecting the limited accuracy of the available navigation information. In comparison, the proposed method produces cleaner warped observations with more regular footprints and reduced boundary artifacts. RDD also removes part of the USBL-related artifacts in some cases, but its behavior is less consistent across all cases. In the latter cases, the seabed relief and texture variations are more complex, making strict visual comparison more difficult. Nevertheless, the warped observations produced by the proposed method still preserve more reasonable covered regions, boundary locations, and footprint orientations than most competing methods. Several baselines exhibit obvious rotation, shear, or displacement deviations, while the raw-scale feature variant becomes less stable when local appearance changes are strong. These observations indicate that the proposed framework improves the reliability of correspondence support, enabling geometrically reasonable alignment on unseen SSS data without retraining.

\section{Conclusion}

This paper proposed SC-Match, a training-free adaptation framework for estimating spatial correspondences between overlapping SSS observations without dense point-level annotations or sonar-specific retraining. The framework adapts pretrained feature extraction and matching components at inference time by combining content-adaptive scale-space feature generation with context-consistent correspondence refinement and alignment. Experiments on two SSS datasets demonstrated improved correspondence accuracy and geometric alignment consistency over representative pretrained matching methods, together with stable alignment behavior under unseen acquisition conditions. The framework has two main limitations. First, the geometric relation is represented by a homography, which captures the dominant transformation but cannot fully model residual non-rigid distortions caused by platform instability, attitude variation, and local seabed relief. Second, correspondences may still occur near shadow trailing boundaries. Although visually salient, these boundaries are view-dependent acoustic effects rather than stable seabed structures, and their locations may vary across survey geometries and track lines. Future work will extend the current homography-based alignment toward spatially continuous deformation modeling and incorporate shadow-aware feature selection or correspondence filtering to improve geometric adaptability and correspondence reliability.




\bibliographystyle{IEEEtran}
\bibliography{refernew}

\end{document}